\documentclass{iopart}             
\usepackage[T1]{fontenc}
\usepackage[applemac]{inputenx}
\usepackage{iopams}
\usepackage{graphicx}
\usepackage{url,graphics,epsfig}
\usepackage{amssymb}
\usepackage[breaklinks=true,colorlinks=true,linkcolor=blue,urlcolor=blue,citecolor=blue]{hyperref}

\usepackage{tikz}

\begin{document}

\jl{2}
%
%
%
\def\etal{{\it et al~}}
\def\newblock{\hskip .11em plus .33em minus .07em}
%
%
%
%
%
%
\setlength{\arraycolsep}{2.5pt}             

\title[Radiative charge transfer for S + H$^+$ collisions]{Radiative charge transfer in cold and ultracold Sulfur atoms colliding with Protons}

\author{G Shen$^{1,2}$, P C Stancil$^{1}\footnote[1]{Corresponding author, E-mail: stancil@physast.uga.edu}$, 
               J G Wang$^{2}$, J F McCann$^{3}$  and B M McLaughlin$^{3,4}\footnote[2]{Corresponding author, E-mail: b.mclaughlin@qub.ac.uk}$}

\address{$^{1}$Department of Physics and Astronomy and  the Center for Simulational Physics,\\
			University of Georgia, Athens, GA 30602-2451, USA}
			
\address{$^{2}$Institute for Applied Physics and Computational Mathematics, Beijing, China}

\address{$^{3}$Centre for Theoretical Atomic, Molecular and Optical Physics (CTAMOP),\\
			School of Mathematics and Physics, The David Bates Building, Queen's University Belfast, 
			Belfast BT7 1NN, UK}

\address{$^{4}$Institute for Theoretical Atomic, Molecular and Optical Physics (ITAMP),\\
			Harvard Smithsonian Center for Astrophysics, MS-14, Cambridge, MA 02138, USA}


%
%

\begin{abstract}
Radiative decay processes at cold and ultra cold
temperatures  for Sulfur atoms colliding with protons are investigated. 
The MOLPRO quantum chemistry suite of codes was used to obtain
accurate potential energies and transition dipole moments, as a function of
internuclear distance, between low-lying states of the SH$^{+}$ molecular cation.
A multi-reference configuration-interaction (MRCI) approximation  together with the Davidson correction is
used to determine the potential energy curves and transition dipole
moments, between the states of interest, where the molecular orbitals (MO's) are obtained from state-averaged
multi-configuration-self-consistent field (MCSCF) calculations. The
collision problem is solved approximately using an optical potential method to obtain radiative loss,
and a fully two-channel quantum approach for radiative charge transfer. Cross sections and rate coefficients are 
determined for the first time for temperatures ranging from 10 $\mu$ K up to 10,000 K. 
Results are obtained for all isotopes of Sulfur, colliding with H$^{+}$ and D$^{+}$ ions and comparison is made to  
a number of other collision systems.
\end{abstract}

%
%


\vspace{1.0cm}
\begin{flushleft}
Short title: Radiative charge transfer in ultracold S and H$^{+}$ collisions\\
\submitto{\jpb: \today}
\end{flushleft}
\maketitle
%

%
%
\section{Introduction}
Molecule formation processes involving second-row elements are of considerable 
interest as searches are ongoing in a variety of interstellar and circumstellar
media \cite{Stancil2000,Benz2005}. The SH$^+$ cation has been observed in 
absorption recently in the diffuse interstellar medium \cite{Menten2011}. 
The growth of state selective ultracold quantum chemistry is 
another motivation for this work \cite{Ospelkaus2010}.
Here we have investigated radiative decay processes at ultracold
temperatures and above for Sulfur atoms colliding with protons. 

Radiative association is the direct combination of two particles, neutral or ionized,
  with de-excitation of the formed molecule by emission of a photon.
 For this process to be efficient it is necessary that the molecule
be formed in a state linked, by permitted transitions, to the fundamental level, so that
it can get rid of its excess energy by emission of a photon.
Previously Stancil et al. \cite{Stancil2000}, studied this system 
  for radiative association in the ground electronic state for S$^+$ ions colliding with atomic hydrogen,
 \begin{equation}
\rm  S^{+} + H \rightarrow SH^{+} + h\nu, 
\label{proc1}
 \end{equation}
while Zhao et al. \cite{zhao2005} performed close-coupling calculations
 of the direct or non-radiative, charge transfer process,
 \begin{equation}
\rm  S + H^{+} \leftrightarrow S^{+}+H. 
\label{proc3}
 \end{equation}
In this work, we use accurate potential energies and transition dipole moments as input to perform collision cross section 
 calculations for the radiative charge transfer process of Sulfur atoms colliding with protons.
 The present collision problem is solved using an optical potential method and a fully quantal approach in order to obtain the radiative
charge transfer cross sections for the process \cite{Stancil2008},
 \begin{equation}
 \rm S(3s^23p^4~^3P) + H^{+} \rightarrow S^{+}(3s^23p^3       ~^4S^{\circ},^2D^{\circ},^2P^{\circ}) + H(1s) + h\nu.
\label{proc2}
 \end{equation}
For excited states of this cation,  for the first time rate coefficients for reaction (\ref{proc2}) 
are determined by averaging over a Maxwellian velocity distribution.

 Rate coefficients are of interest for theoretical models of photodissociation regions (PDRs) 
 and X-ray-dominated regions (XDRs) as  S$^{+}$ ion-chemistry
may be important \cite{Sternberg1997}. S$^{+}$ ion-chemistry may 
also play a role in the X-ray chemistry in the envelopes 
 around young stellar objects \cite{Benz2005}. 
 For the similar ion CH$^+$ the rate coefficient for its formation via 
 radiative association ranges between 10$^{-14}$ and 10$^{-13}$cm$^3$s$^{-1}$. 
  The formation rates for SH$^+$ are expected  to be comparable.
Here we determine rate coefficients for temperatures ranging from 10 $\mu$ K up
to 10,000 K for these applications and also investigate isotope effects. 
Our results are compared to a number of other ion-atom collision systems.

The layout of this paper is as follows.  In section 2 we give a brief outline of the theoretical approaches used 
to determine the cross sections and rate coefficients. Section  3 presents the results from our work and 
finally in section 4 conclusions are drawn from our present investigations. Atomic units are used throughout unless otherwise noted. 
 
\section{Theory}\label{sec:Theory}
 
 \subsection{Electronic structure calculations}
Following our earlier work on this molecular cation \cite{Stancil2000} for radiative association,
 we extend our computations using a parallel version of the 
 MOLPRO~\cite{Werner2010} suite of {\it ab initio} quantum chemistry codes 
(release MOLPRO 2010.1) to calculate the molecular structure of this diatomic hydride SH$^+$ 
to higher-lying triplet electronic states and the transition dipole matrix elements connecting the  states.
Potential energy curves and transition dipole moments as a function 
of internuclear distance are computed out to a bond separation of  $R=18$ a.u.  For separations beyond this, 
we use a multipole expansion to represent the long-range part of the potentials. 
We conducted Multi-Reference Configuration Interaction (MRCI)
calculations on the State-Averaged Multi-Configuration-Self-Consistent-Field 
(SA-MCSCF) wavefunctions \cite{Helgaker2000}. 
The Davidson  correction was applied to all our results \cite{Davidson1974}.
This multi-reference configuration interaction (MRCI) approach with the Davidson correction 
was used to calculate all the potentials for this molecular cation as it dissociates.
To model the all-electron calculations, we use an augmented-correlation-consistent 
polarized valence sextuplet  Gaussian basis set; aug-cc-pV6Z (AV6Z). The choice of this AV6Z basis 
is due to the fact that in quantum chemistry calculations these large Gaussian basis sets  
are well known to recover $\sim$ 99 \% of the electron correlation energy \cite{Helgaker2000}.
All our electronic structure computations were performed in the C$_{2v}$ 
Abelian point group symmetry (A$_1$, B$_1$, B$_2$, A$_2$). 
	%
	%
\begin{figure}
\includegraphics[width=\textwidth]{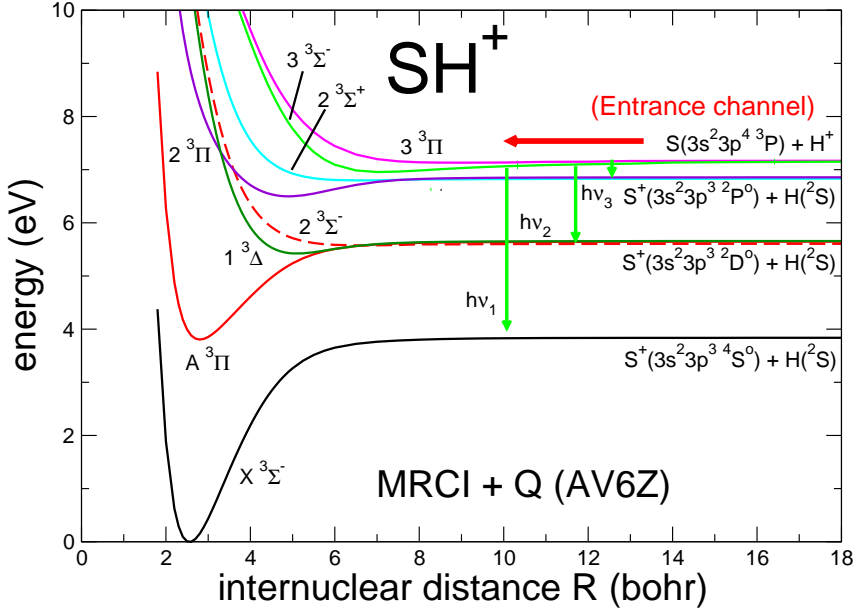}
\caption{(Colour online) Relative electronic energies (eV) for the ${\rm SH}^+$ molecular cation as a 
               function of bond separation at the MRCI+Q level of approximation with an AV6Z basis. 
               The 3 $\rm ^3\Sigma^{-}$  and 3 $\rm ^3\Pi$ entrance channels of the 
               ${\rm S (3s^23p^4~ ^3P}) + {\rm H^{+}}$ cation are shown together with lower lying triplet electronic states 
               ($\rm ^3\Sigma^{-}$, $\rm ^3\Sigma^{+}$,  $\rm^3\Pi$  and  $\rm ^3\Delta$) for which 
               radiative charge transfer may occur.
               The allowed radiative transitions from the entrance channels equation (3)
               are indicated by the vertical lines.}
\label{fig1}
\end{figure}
 	%
 	%
\begin{figure}
\includegraphics[width=\textwidth]{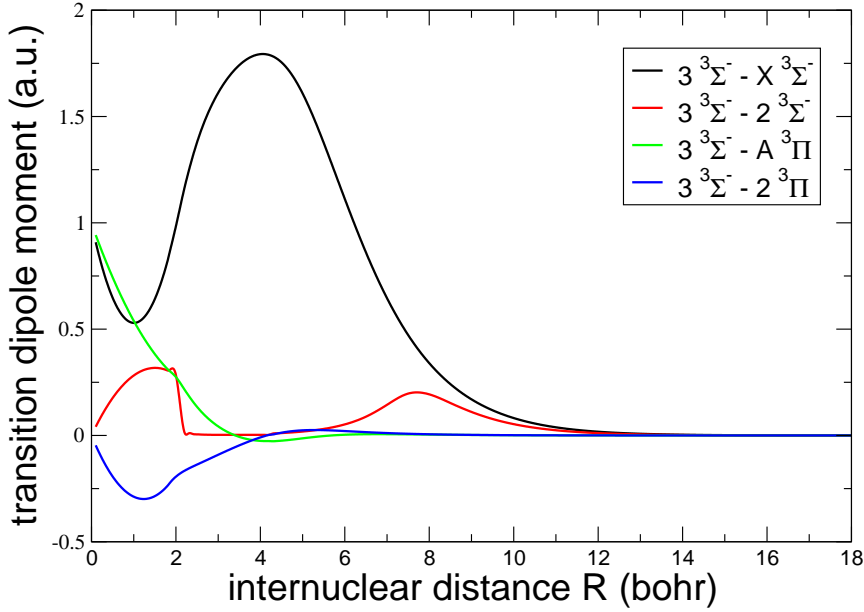}
\caption{(Colour online) Dipole transition moments $D(R)$ for the ${\rm 3}~^{3}\Sigma^{-} \rightarrow
 		{\rm X}~ ^{3}\Sigma^{-},  2~^{3}\Sigma^{-},  2~^{3}\Pi, A~^{3}\Pi$ transitions.  The  MRCI + Q
		approximation with an AV6Z basis set  
		 is used to calculate the transition dipole moments.}
\label{fig2}
\end{figure}
  	%
	 %
\begin{figure}
\includegraphics[width=\textwidth]{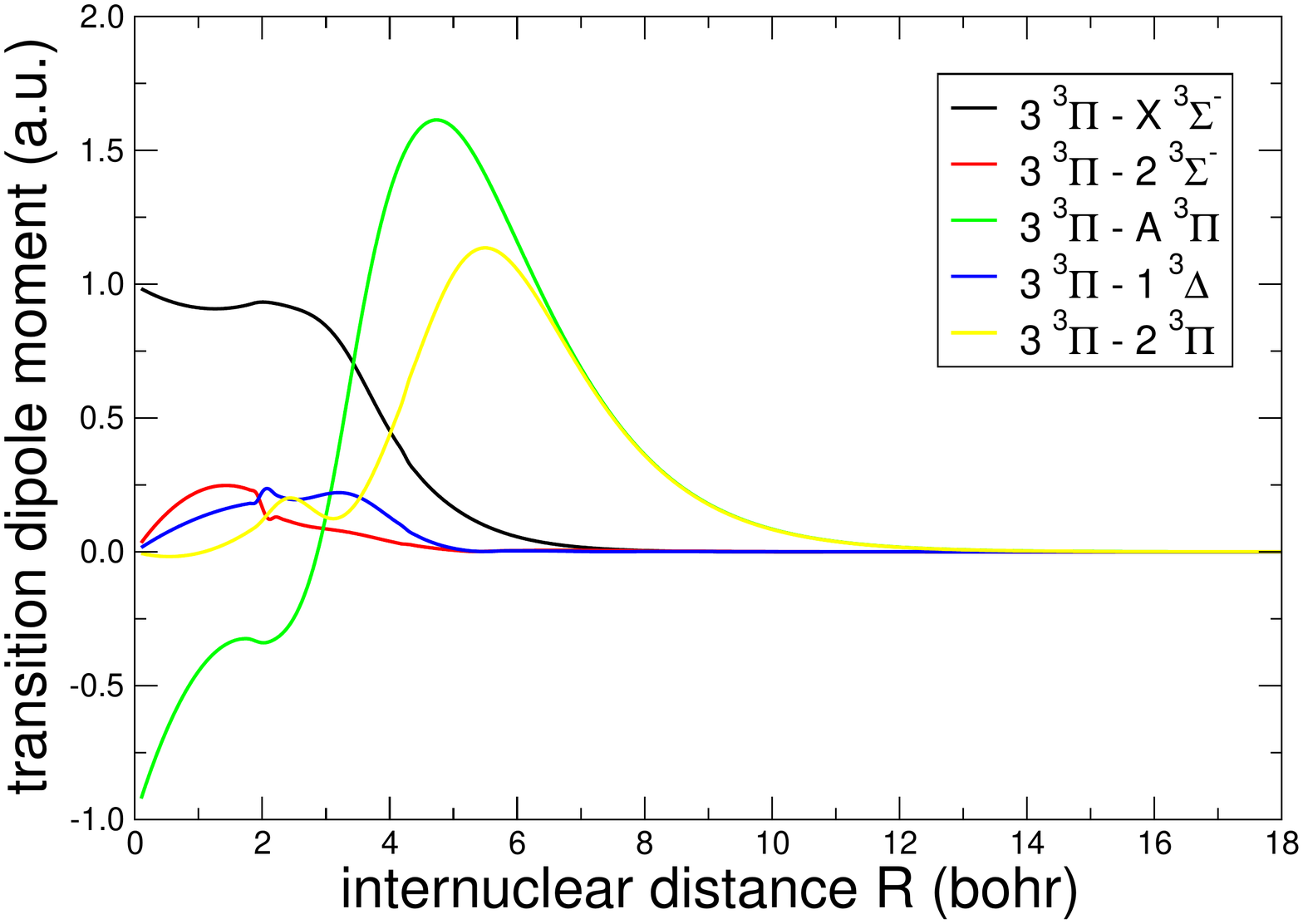}
\caption{(Colour online) Dipole transition moments $D(R)$ (in atomic units) 
		for the ${\rm 3}~^{3}\Pi \rightarrow
 		{\rm X} ~^{3}\Sigma^{-}, 2~^{3}\Sigma^{-}, A~^{3}\Pi, 1~^{3}\Delta, 2~^{3}\Pi$ transitions. 
		The MRCI+Q approximations with an AV6Z basis set 
		 is used to calculate the transition dipole moments.}
\label{fig3}
\end{figure}
In the  C$_{2v}$  point group, all molecular orbitals were labeled by their symmetry ($a_1$,$b_1$,$b_2$,$a_2$); when
symmetry is reduced from C$_{\infty v}$ to C$_{2v}$, the correlating relationships
are $\sigma \rightarrow a_1$, $\pi \rightarrow$ ($b_1$, $b_2$) , $\delta \rightarrow$ ($a_1$, $a_2$). 
The active space consists of 16 electrons and 10 molecular orbitals $6a_1$, $2b_1$, $2b_2$, $0a_2$ (6 2 2 0).
To take account of short-range interactions  we employed the multi-configuration-self-consistent-field (MCSCF) 
method~\cite{Werner1985,Knowles1985} available within the MOLPRO~\cite{Werner2010} suite of codes.  
The molecular orbitals for the MRCI method were obtained from the 
state-averaged-multi-configuration-self-consistent- field method (SA-MCSF). 
The averaging processes was carried out on the lowest four $^3\Pi$, 
three $^3\Sigma^-$ and three $^3\Delta$ molecular states of this cation.

In Fig. \ref{fig1} we illustrate all the triplet states involved in the radiative decay processes.
 All the potential energies in Fig. \ref{fig1} are given in eV relative to the ground-state
equilibrium bond distance $r_e$ of this cation.  Note, radiative charge transfer can occur either from  
the 3 $\rm ^3\Sigma^-$ or the  3 $^3\Pi$ state (the entrance channels).
 The 3 $\rm ^3\Sigma^-$ has a shallow well whereas the 3 $^3\Pi$ state is totally repulsive.
Table 1 gives asymptotic properties of the molecular states.

Figs. \ref{fig2} illustrates the dipole transition moments $D(R)$  as a function of internuclear separation $R$ 
connecting the 3 $\rm ^3\Sigma^-$  to the lower lying triplet  electronic states involved in the dynamics. 
In Fig. \ref{fig3} the corresponding transition dipole moments connecting the 3 $^3\Pi$ entrance channel are shown.
The results  in Figs. \ref{fig2} and \ref{fig3} are obtained from the MRCI+Q calculations 
using an AV6Z basis.
 We note that the asymptotic separated-atom  energies in the MRCI+Q approximation 
 show suitable agreement with experimental atomic values (see Table 1).

 Our results show smooth  dipole moments connecting all of the adiabatic 
 triplet states which leads us to conclude that the sensitivity 
of the radiative transitions will be primarily due to the nuclei wave function envelope. 
The resonance behaviour will be primarily due to potential scattering in the elastic scattering channel.

 The spontaneous decay rate $\Gamma (R)$ (see below) decreases exponentially as  $R$ increases due 
to the exponential attenuation in the overlap of the atomic wave functions corresponding to charge transfer. 
Beyond $R$ = 18 a.u., the potential of the  ${\rm 3}~ ^{3}\Sigma^{-}$ or ${\rm 3}~^{3}\Pi$ excited state 
(the entrance channel, see Fig. \ref{fig1}) can be approximated by the long-range multipole expansion:
\begin{equation}
 V(R) = \frac{Q_M}{2R^{\mathrm{3}}}-\frac{\alpha_d}{2R^{\mathrm{4}}}	+	V(+\infty),
\label{mult}
\end{equation}
where $Q_M$ is the quadrupole moment and $\alpha_{d}$ is the  dipole polarizability of the neutral atom.
$Q_M$ for a $p^4$ configuration is $-\frac{4}{5}e<r^2>$ for a $\Sigma$ state and $\frac{2}{5}e<r^2>$ for a $\Pi$ state \cite{Gentry1977}, 
where the $<r^2>$ value for Sulfur is 5.065255 a.u ($^3P$ state) \cite{Fisher1972,Fisher1973} and $Q_M=0$ for H.
The atomic values for the dipole polarizability adopted here are respectively, $\alpha$(S)=18.0 and  $\alpha$(H)=4.5.  
The long-range expansion of the final channel has just the last 2 terms of the right hand side in (\ref{mult}). 
At short range internuclear distances we fitted the potentials with the form, 
\begin{equation}
 V = A e^{-BR} +  C
\end{equation}
where $A$, $B$ and $C$ are fitting constants. A similar approach was used for 
the transition dipole moments to extend to long 
and short range internuclear distances for our cross section calculations.
\begin{table*}[!ht]
\caption{Asymptotic Separated-atom and United-atom limit Properties of the molecular states included in our present work.}
\begin{tabular}{llcccc}
\hline
\multicolumn{1}{l}{Molecular} & \multicolumn{3}{c}{Separated-atom} &
\multicolumn{2}{c}{United-atom  Cl$^{+} $} \\
\cline{2-4}
\cline{5-6}
\multicolumn{1}{l}{State} & \multicolumn{1}{c}{Atomic states} &
\multicolumn{2}{c}{Energy (eV)} & 
\multicolumn{1}{c}{State} &
\multicolumn{1}{c}{Dipole Moment$^c$} \\
\cline{3-4}
\multicolumn{2}{c}{} & \multicolumn{1}{c}{Theory$^a$} &
\multicolumn{1}{c}{Expt.$^b$} & \multicolumn{2}{c}{(a.u.)} \\
\hline
$3~^3\Sigma^-$ & S($3s^23p^4~^3P$) + H$^+(^1S_g)$ &\, 0.00 &\, 0.00 & $3s^23p^3(^4S^o)3d~^3D^o$ & -\\

$2~^3\Pi$ & S$^+$($3s^23p^3~^2P^o$) + H$(^2S_{\frac{1}{2}})$ & -0.31 & -0.22  & $3s3p^5~^3P^o$ &\, 0.0\\

$A~^3\Pi$ & S$^+$($3s^23p^3~^2D^o$) + H$(^2S_{\frac{1}{2}})$ & -1.52 & -1.42  & $3s^23p^4~^3P$ & \,0.997\\

$2~^3\Sigma^-$ & S$^+$($3s^23p^3~^2D^o$) + H$(^2S_{\frac{1}{2}})$ & -1.51 & -1.42  & $3s^23p^3(^4S^o)4s~^3S^o$ & \,0.0\\

$X~^3\Sigma^-$ & S$^+$($3s^23p^3~^4S^o$) + H$(^2S_{\frac{1}{2}})$ & -3.32 & -3.26  & $3s^23p^4~^3P$ &\, 0.997\\
\hline 
$3~^3\Pi$ & S($3s^23p^4~^3P$) + H$^+(^1S_g)$ & -0.00 &\, 0.00  & $3s^23p^3(^4S^o)3d~^3D^o$ & -\\

$2~^3\Pi$ & S$^+$($3s^23p^3~^2P^o$) + H$(^2S_{\frac{1}{2}})$ & -0.31 & -0.22  & $3s3p^5~^3P^o$ &\, 0.0\\


$1~^3\Delta$ & S$^+$($3s^23p^3~^2D^o$) + H$(^2S_{\frac{1}{2}})$ & -1.51 & -1.42 & $3s^23p^3(^4S^o)3d~^3D^o$ &\, 0.0\\

$A~^3\Pi$ & S$^+$($3s^23p^3~^2D^o$) + H$(^2S_{\frac{1}{2}})$ & -1.52 & -1.42 & $3s^23p^4~^3P$ & -0.997\\

$2~^3\Sigma^-$ & S$^+$($3s^23p^3~^2D^o$) + H$(^2S_{\frac{1}{2}})$ & -1.51 & -1.42 & $3s^23p^3(^4S^o)4s~^3S^o$ &\, 0.0\\

$X~^3\Sigma^-$ & S$^+$($3s^23p^3~^4S^o$) + H$(^2S_{\frac{1}{2}})$ & -3.33 & -3.26 & $3s^23p^4~^3P$ &\, 0.997\\
\hline 
\end{tabular}
\\
$^a$MRCI+Q (AV6Z basis).\\
$^b$Deduced from NIST Atomic Spectra Database tabulations \cite{NIST2012}.\\
$^c$Transition dipole moments with initial states, deduced from \cite{Tayal2004}.
\end{table*}
		%
		%
%
%
%
%
%
\subsection{Dynamics}
In the simple classical scattering model \cite{Miller70}, the nuclear motion takes place on the ion-atom 
incoming potential surface, $V(R)$. Thus in a first approximation the motion is  angular-momentum 
conserving, time-reversal invariant, and elastic. 
With the centre-of-mass frame denoted  as $E$ 
and the reduced mass of the nuclei  as $\mu$,  then we  take,  
$V(+\infty) =0$.  So for  an impact parameter, $b$, the semi-classical cross-section is simply,
\begin{equation}
\sigma_c  (E)  = 2\pi \int_0^{+\infty} b P(E,b) db, 
\end{equation}
where $P(E,b)$ is the capture probability.

Since the radial velocity can be 
written as,
\begin{equation}
v_{R}^{2}(R) =  {2 E \over  \mu}  \left(  1 - { V(R) \over E} -{ b^2 \over R^2}  \right),
\end{equation} 
then the classical turning point will be the (largest) solution of the equation:
\begin{equation}
{dR(t) \over dt} = v_{R}(R_c)=0.
\end{equation} 
The process of spontaneous emission has a rate $\Gamma(R)$ which drives 
the charge transfer process.  
One can write for the probability of emission, for example as explained in  \cite{Miller70}, 
for  weak coupling as:
\begin{equation}
P(b,E) \approx  2 \int_{R_c}^{+\infty} { \Gamma (R)  \over v_{R}(R)} dR,   
\end{equation}
which leads directly to the expression, 
\begin{equation}
 \sigma_c (E) =  2 \pi \sqrt{\frac{2\mu}{E}} \int_0^{+\infty} b~ db \int_{R_c}^{\infty} \frac{ \Gamma (R) \ dR}{\sqrt{1 - V(R)/E - b^2/R^2}}.
 \label{semic}
 \end{equation}
 At high energies,  $E \gg V$, the trajectory is almost rectilinear and
  the integrand is energy independent  and thus $\sigma (E) \sim (\mu/E)^{1/2}$.
  It is purely by coincidence  that this  energy dependence, which arises from the collision time,   
  matches the  classical Langevin model \cite{lang05} for reactive collisions.
  	%
	%
\begin{figure}
\includegraphics[width=\textwidth]{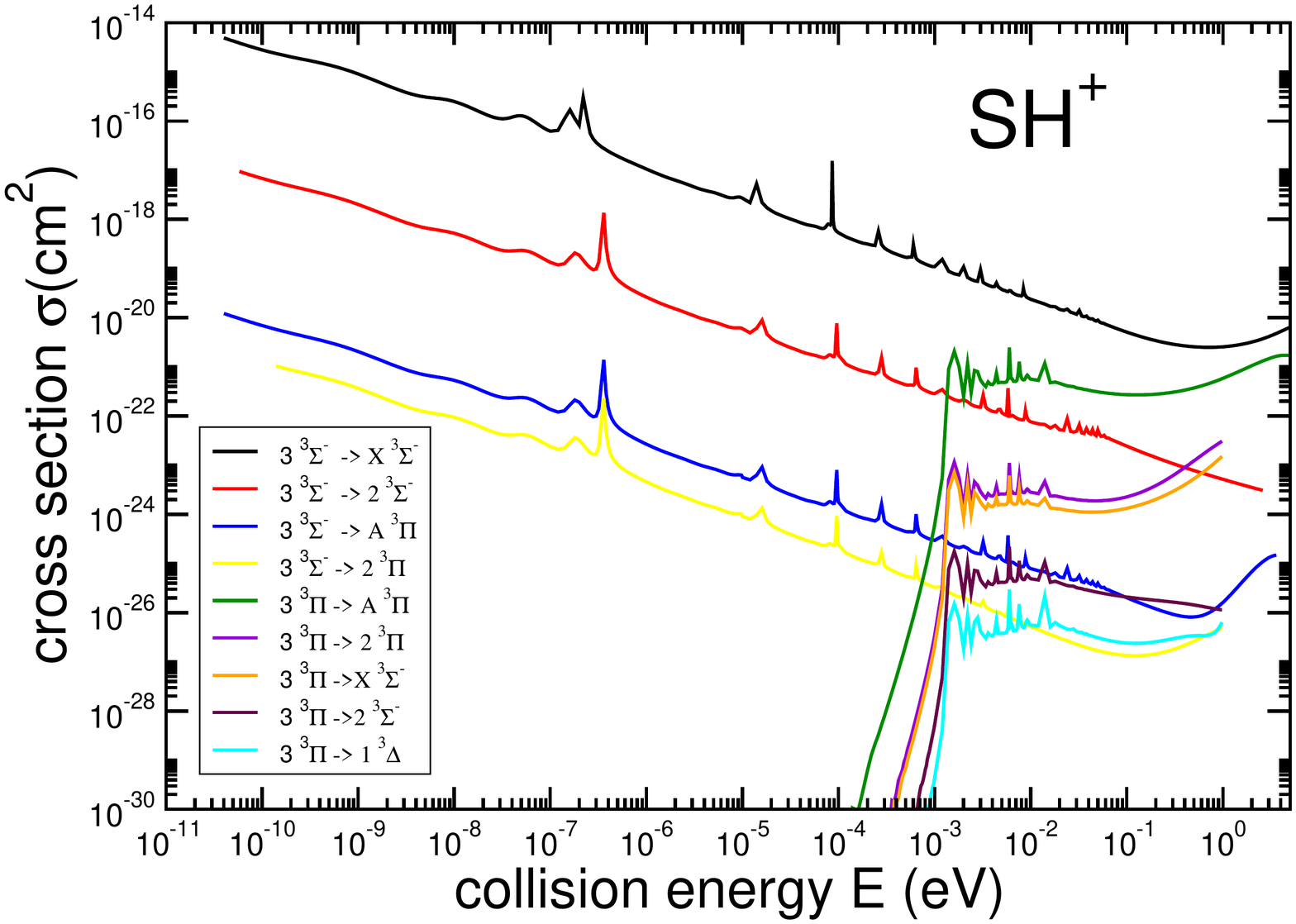}
\caption{(Colour online) Cross section $\sigma$ (cm$^2$) for radiative charge transfer as a 
	       function of centre of mass collision energy $E$ (eV), for the nine allowed transitions
	       obtained with the optical potential method, see text for details. 
	       In the figure we present the spinless cross sections,
		that is g = 1.}
\label{sh+rcx}
\end{figure}
  	%
	%
\begin{figure}
\includegraphics[width=\textwidth]{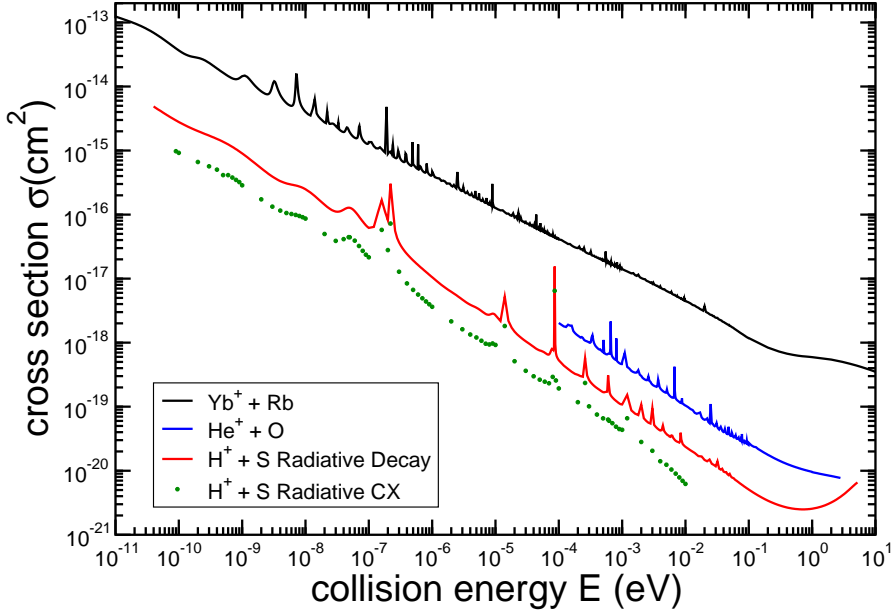}
\caption{(Colour online) Total radiative charge transfer cross sections  for H$^{+}$ + S, 
		He$^{+}$ + O \cite{Zhao2004} and Yb$^{+}$ + Rb \cite{McLaughlin2014}.
		Cross sections $\sigma$ are in (cm$^2$) as a function of centre of mass collision energy $E$ (eV), 
		for the various systems, see text for details.}
\label{rcx-ions}
\end{figure}
  	%
 	 %
\begin{figure}
\includegraphics[width=\textwidth]{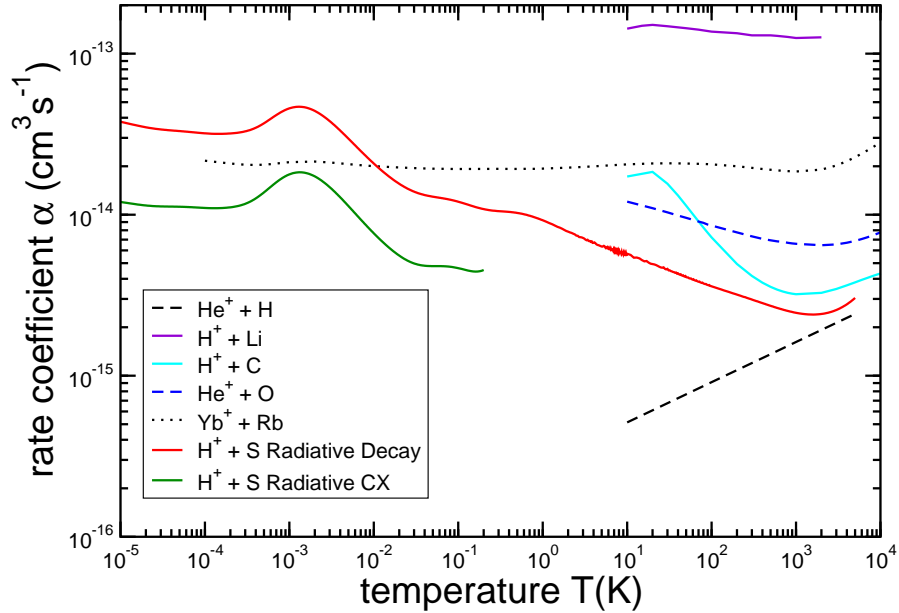}
\caption{(Colour online) Rate coefficients $\alpha$ (cm$^3$s$^{-1}$) as a function of temperature $T$ (K) 
		for the various molecular cations obtained from a Maxwellian average
	     	of the radiative charge transfer cross sections, see text for details.} 
\label{rate-ions}
\end{figure}
  	%
	%
\begin{figure}
\includegraphics[width=\textwidth]{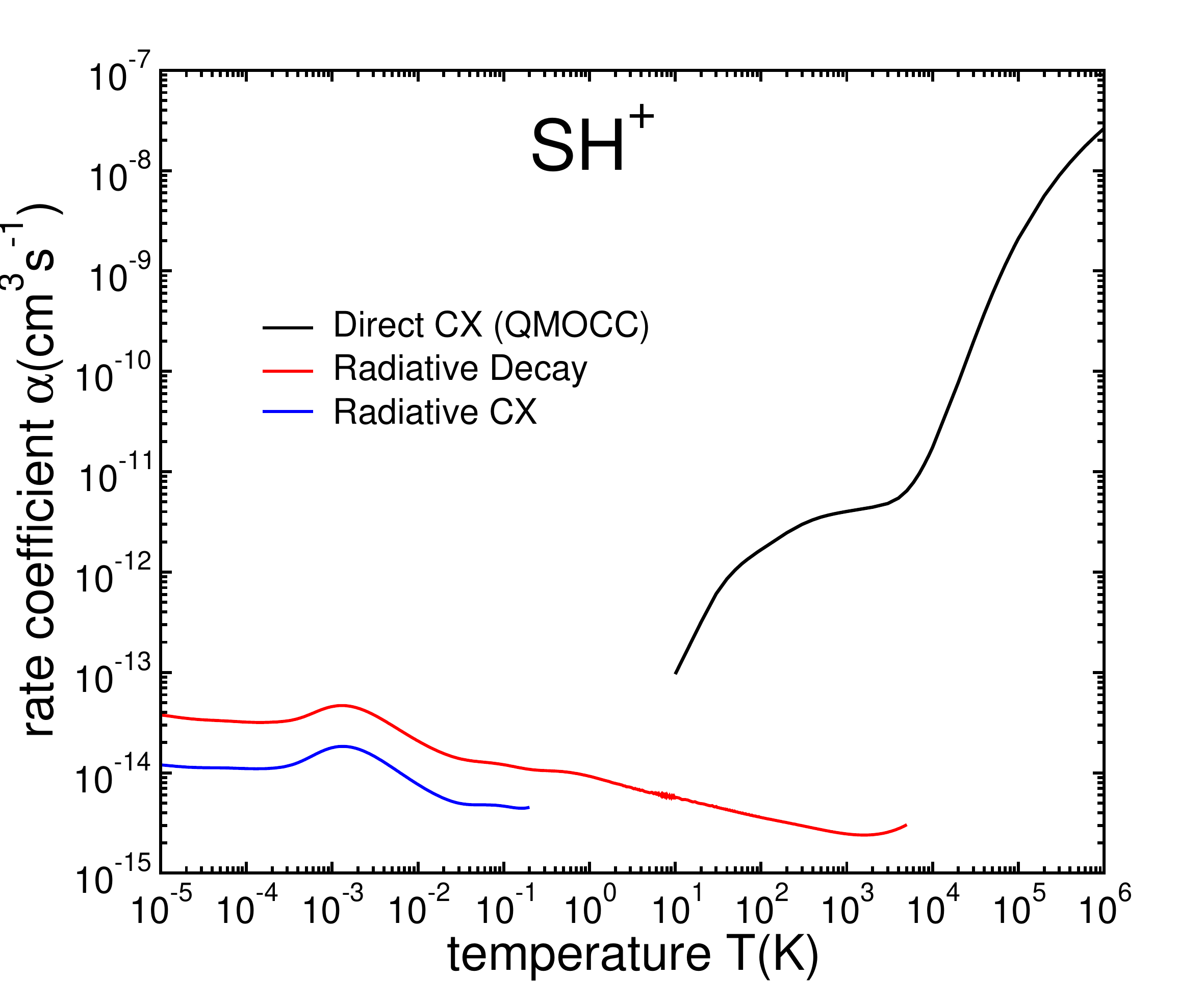}
\caption{(Colour online) Rate coefficients $\alpha$ (cm$^3$s$^{-1}$) as a function of temperature $T$(K) 
		for S  + H$^{+}$ obtained from a Maxwellian average of the radiative decay cross sections. 
		Results are shown for both the radiative decay (red line), radiative charge transfer (blue line)
	    	and a fully Quantal-Molecular-Orbital-Close-Coupling (QMOCC, black line) \cite{zhao2005}
	      	for direct charge exchange, see text for details.} 
\label{sh+rate}
\end{figure}
 The optical potential method, in the context of radiative charge transfer, 
 has been described in detail by Dalgarno and co-workers \cite{Cooper1984,zygel88,zygel89}.
  In the adiabatic approximation the dynamics occur on decoupled, centrally-symmetric potential energy curves.  
In the ultracold regime ($E$ <  1 meV) all  non-adiabatic radial and rotational coupling are weak and  the 
vacuum coupling (by photo-emission) is the dominant  non-elastic process. The ion-atom collision 
channel thus becomes a resonance state, with an effective potential 
associated with the radiative shift and width.  
The physics is thus reduced to an optical potential: 
 single channel (complex  central potential) scattering problem:
 \begin{equation}
 V_c(R) = V(R)- \textstyle{1 \over 2} i~ \Gamma (R),
 \end{equation} 
 where $V(R)$ is the real (shifted) adiabatic potential in the ion-atom collision potential.
 
In this paper, the  imaginary part of optical potential method represents loss though radiative 
emission, which include decay to a bound or continuum state of the molecule ion. That is, implicitly, and approximately, 
we include  both the process of radiative association and radiative charge transfer. This point is discussed 
in detail in recent applications \cite{zyge14,sayf13} and its validity verified. In other terms,  
the Einstein spontaneous decay rate $\Gamma$, which  
is larger the higher the photon frequency,  is taken as a vertical transition in analogy to  the 
way that the `reflection principle'  is applied \cite{schi93}. 
This approximation becomes more accurate the heavier the mass of the colliding atoms/ions, 
since the momentum (local wavelength) of the nuclei  is conserved when the reduced mass is larger. 
The problem can be summarized mathematically  \cite{zyge14}   as the solution of the 
 Schr\"odinger equation,
\begin{equation}
\left[ -\frac{1}{2\mu}{{\mathop{\nabla}}^2_{\bf R}} + V (R) - E \right]  F(E; {\bf R}) = \frac{1}{2} i~ \Gamma (R) F(E; {\bf R}), 
\end{equation}
where,
\begin{equation}
\Gamma (R) = \frac{4 D^2 (R)}{3c^{3}}  \left| V(R) - V_f (R) \right | ^3,
\label{trate}
\end{equation}
and, using atomic units, $c$ is the speed of light, where $V(R)$ and  $V_f (R)$ are the  adiabatic potentials of the  
upper (initial) and lower (final) electronic state. Since the overlap of the electronic states involved in the charge transfer 
 is large only at short-range, $\Gamma(R)$ is  exponentially damped with increasing $R$. 
  As the potential is central, even though it is complex, the usual separation in spherical coordinates applies, 
  and  the scattering wavefunction can be decomposed as,
 \begin{equation}
 F(E ; {\bf R})  = \sum_{J,M_J} \chi_{J} (k, R) Y_{JM_J} (\hat{\bf R}), 
 \label{sph}
 \end{equation}
 where $Y_{JM_J}$ and $\chi_{J}$ are the spherical harmonics and radial wave function, respectively.
 We define the elastic-scattering wavenumber, $k_{J}(R)$,  for the initial ion-atom channel 
 with angular momentum $J$, as follows:
 \begin{equation}
 k_{J}^2(R) = k^2 -2\mu  V(R) -J(J+1)/R^2. 
 \end{equation}
 The collision wavenumber is defined as
 \begin{equation}
 k = \lim_{R \rightarrow \infty}  k_{J} 
 \label{wavk}
 \end{equation}
 and the corresponding radial functions,  $f_{u,J}(k, R)=kR \chi_{u,J} (k, R)$, will be the solutions of the equations:
 \begin{equation}
\left[ \frac{d^2}{dR^2}+k_{J}^2(R)   \right] f_{J}(k, R) = 0. 
\label{eqn9}
 \end{equation}
This is normalized asymptotically ($R \rightarrow \infty$) according to,
 \begin{equation}
 f_{J}(k,R) \rightarrow  \sqrt {\frac{2 \mu}{\pi k}}\sin  \left(  k R - \frac{1}{2}J\pi + \delta_J\right)
 \label{normc} 
 \end{equation}
and $\delta_J$ is the elastic  phase shift.  When the optical potential is used 
 the radial equations for the functions in (\ref{sph}) are the same:
 \begin{equation}
 \left[ {d^2 \over dR^2} +\kappa^2_{J}(R)   \right] f_{cJ} (k,R) =0,
 \label{chiq}
 \end{equation}
 apart from the modification for the  complex wavenumber:
 \begin{equation}
\kappa^2_{J}(R) = k_{J}^2(R) -i\mu \Gamma(R), 
 \end{equation}
 and the corresponding complex radial function, $f_{cJ}$.
 
The imaginary term is short-ranged, so the asymptotic wavenumber, and hence  
the normalisation conventions (and density of states) for the radial wavefunctions  (\ref{normc}) are the same.  However, 
the partial waves,    $f_{cJ}(k,R)$, have  complex phase shifts  \cite{mott65} and thus the outgoing  probability flux is  attenuated. 
     	%
	%
\begin{figure}
\includegraphics[width=\textwidth]{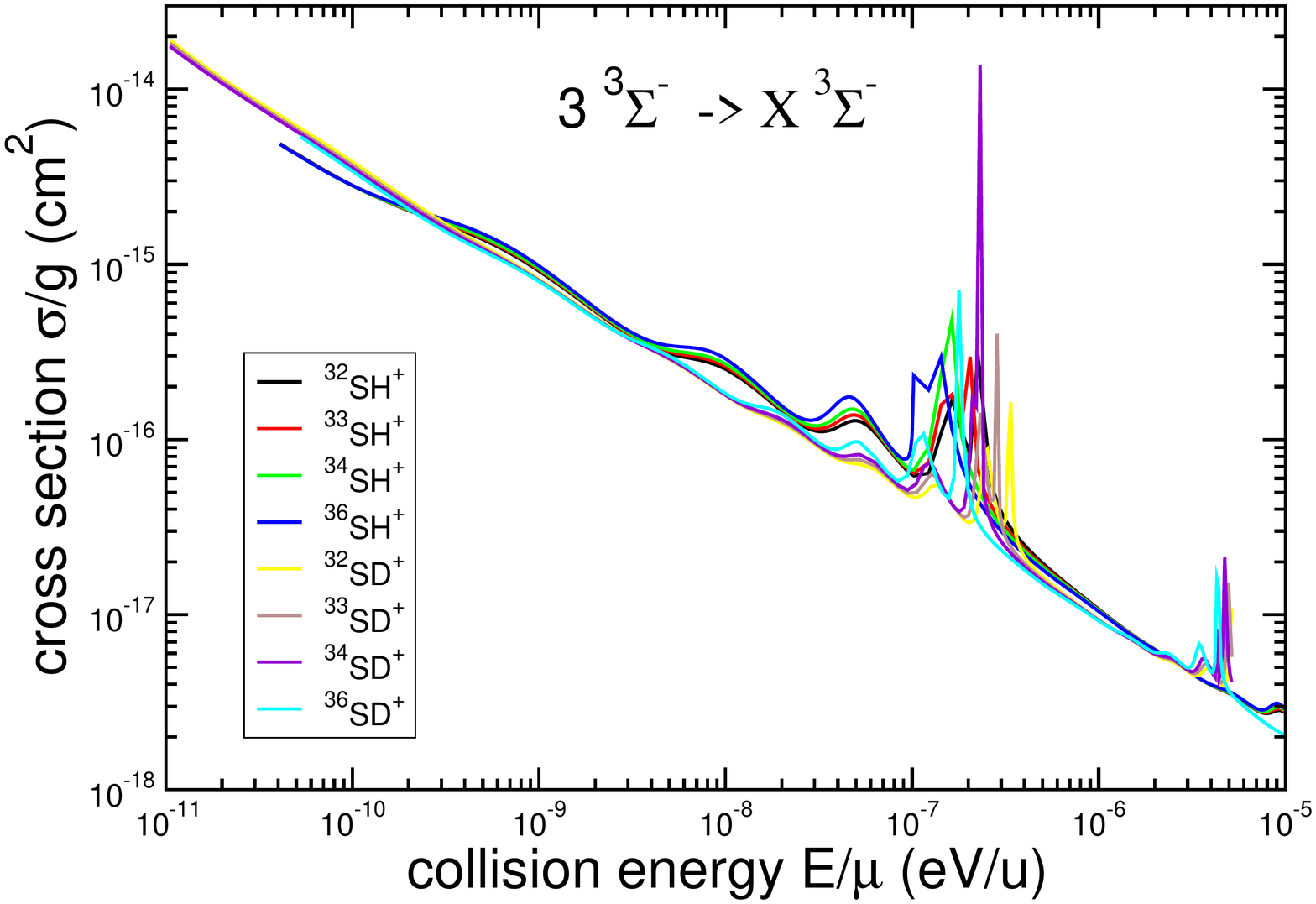}
\caption{\label{isotope1}(Colour online) Radiative charge transfer cross section $\sigma$ (cm$^2$), for all the various isotopes of Sulfur 
		colliding with H$^+$ and D$^+$ ions, as a function of the centre of mass collision energy $E$ (eV) per reduced mass $\mu$, 
		for the dominant ${\rm 3}~^{3}\Sigma^{-} \rightarrow {\rm X}~ ^{3}\Sigma^{-}$ transition, see text for details.}

  	%
	%
\end{figure}
\begin{figure}
\includegraphics[width=\textwidth]{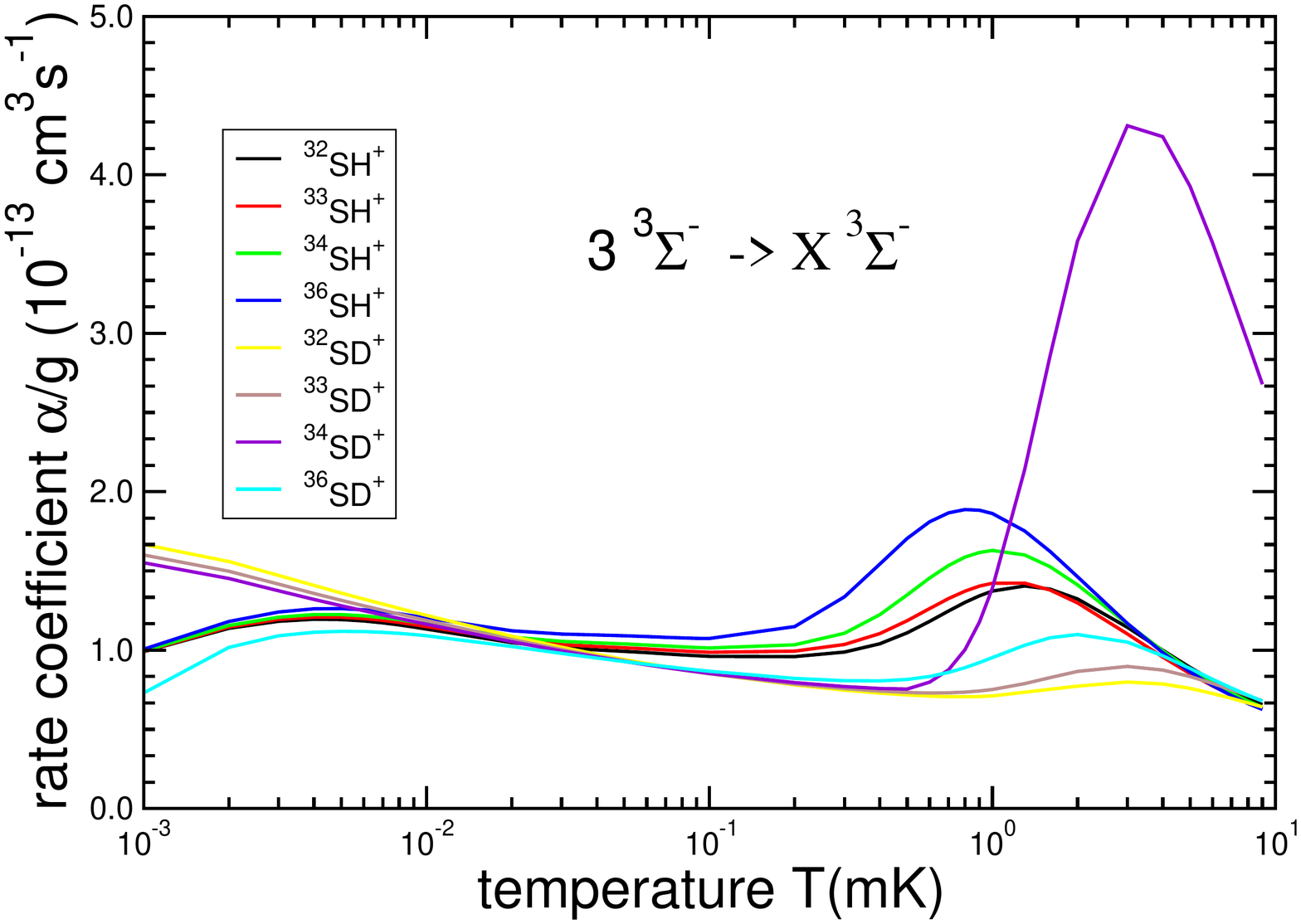}
\caption{\label{isotope2}(Colour online) Radiative charge transfer rate coefficient $\alpha$ (10$^{-13}$ cm$^3$s$^{-1}$) for all the 
		various isotopes of Sulfur colliding with H$^+$ and D$^+$ ions 
		as a function of the temperature $T$ (mK), for the dominant 
		${\rm 3}~^{3}\Sigma^{-} \rightarrow {\rm X}~ ^{3}\Sigma^{-}$ transition, see text for details.}

\end{figure} 

  Naturally  the vacuum emission represented by the width $\Gamma (R)$ is much smaller in magnitude 
  compared with the real potential $V(R)$ and thus we can solve (\ref{chiq})  by perturbation theory. 
 In the distorted-wave approximation  the imaginary part of the phase shift, $\mu^{DW}_J$,  is given by
 \begin{equation}
 \mu^{DW}_J(k) = \frac{\pi}{2} \int_{0}^{+\infty} | f_{J}(k, R)| ^ 2 \Gamma (R) dR.
 \label{dwa}
 \end{equation}
Here the collision problem is solved by directly  integrating equation (\ref{chiq})  \cite{alli70,alli72,john77}. 
To distinguish this from the above discussed semiclassical and distorted-wave approximations, we
refer to this below as the {\em quantal approximation}. 

As regards the elastic scattering cross section. This is much more problematic. In the ultra cold regime, say 
sub meV, where one can apply a modified version of effective range theory in which  the elastic scattering 
(real part of the phase shift) can be described by the scattering length. 
The  real phase shift, is extremely difficult to calculate at  threshold. The sign and magnitude of the 
scattering length, connected to the energy of the last bound state below threshold of the well, 
determines the cross section. Even a small energy shift or correction to the long-rang part of the potential, 
such as  hyperfine splitting  makes an important contribution, and of course relativistic effects play a role also. 
However, for the reactive transition described in this paper, the imaginary part of the phase shift, only depends on the 
short range overlap of the atomic orbitals where the de Broglie wavelength  is short, and thus is not  sensitive to  
long-range correlations and corrections. Thus we can be confident in the robustness  of our results 
to such small corrections.

 The cross section for total collision-induced radiative decay from the excited state entrance channel, 
 the sum of the cross sections for processes (\ref{proc2}) and the process,
\begin{equation}
\rm  S + H^{+} \rightarrow SH^{+}(^{\ast}) + h\nu \label{proc3}, 
\end{equation}
can be obtained within the optical  or quantal approximation.
 The cross section for collision-induced radiative decay can then be written as,
\begin{equation}
\sigma (E) = \frac{g\pi}{k^2} \sum_{J=0}^{\infty} (2J + 1){\left[ 1 -  e^{-4 \mu_J} \right]},
\label{quantal}
\end{equation}
where $k$ is given by (\ref{wavk}), and $g$ is the statistical weight or approach probability factor, 
$g=1/3$ for the ${\rm 3}~^{3}\Sigma^{-}$ state and 2/3 for the  ${\rm 3}~^{3}\Pi$.
At higher energy,  a semi-classical approximation may be invoked to calculate the cross sections for radiative
decay.  The summation over the angular momentum in equation (\ref{quantal}) can be replaced by an integral over the 
impact parameter, $b$, according to $ k b \approx J$. 
The JWKB approximation can then be used to obtain the elastic scattering wave function, 
\begin{equation}
f_{J}(k, R) \approx  \sqrt {\frac{2 \mu}{\pi k_{u,J} (R)}}\sin  \left( \int_{R_c}^R k_{u,J}(R') dR' +\textstyle{ 1 \over 4} \pi \right).
\end{equation}
This  simplifies the calculation of the phase-shift equation (\ref{dwa})
  \cite{zygel88,alli66,bates51}  since the rapidly varying integrand gives us 
  (in the classically allowed region)
$ \vert f_{J}(k, R) \vert^2 \approx   \mu/(\pi k_{J}(R))$. Then 
using equation (\ref{dwa}) we obtain  the semi-classical approximation (\ref{semic}).

The velocity averaged rate coefficient $\alpha  =\langle v \sigma \rangle $, 
 as a function of temperature $T$ (in Kelvin), is obtained by integrating 
  over the Maxwellian distribution \cite{bates51}. That is the rate coefficient is given by,
 \begin{equation}
 \alpha (T) = \left( { 8 \over \mu \pi k_{B}^{3} T^{3}} \right)^{1/2}  \int_{0}^{\infty} E ~ \sigma (E) e^{- E/(k_B T)}  d E.
\label{rate}
 \end{equation}
 We use this expression to evaluate our rates rather than define an effective energy
 depend rate $v\sigma$ as is normally used in cold collisions  \cite{Krems2010,McLaughlin2014}. 
 
\section{Results and Discussion}\label{sec:Results}

Cross sections were obtained using  the optical potential method with the 
SH$^{+}$ potential energies and transition dipole moments, 
obtained within the MRCI+Q  approximation.        
Upper limits to the radiative charge transfer cross sections were 
calculated and are presented  in Fig \ref{sh+rcx}. 
The potentials were shifted to match the experimental asymptotic energies. 
This shift is quite small as can be seen from Table 1 where comparisons 
between theory and experiment are presented.	 

The dominant transition as can be seen from Figs. \ref{fig2} and \ref{fig3} 
correspond to the $\rm3~ ^{3}\Sigma^{-}  \rightarrow  X~ ^{3}\Sigma^{-}$, 
while all transitions which originate from the $\rm 3~^3\Sigma^{-}$ electronic state of SH$^{+}$ 
have a Langevin $1/v$ or $E^{-1/2}$ dependence at energies below 
$\sim10^{-7}$ eV. Since the  $\rm 3 ~^3\Pi$ electronic state is completely repulsive except 
a small potential barrier at large internuclear distance, transitions which originate with this
electronic state are quite small at energies below $\sim10^{-3}$ eV, 
but become competitive above $\sim$ 1 eV as they
have the largest transition dipole moments (See Fig. \ref{fig3}). 
Since the $\rm 3~ ^{3}\Sigma^{-}$ electronic state of SH$^{+}$ is slightly attractive, 
it can therefore support a number of quasi-bound states which give rise to the large
number of  resonances superimposed on the background cross sections.
The energy dependence of the background cross section follows, to a very good approximation,
that derived from the simple semiclassical argument.  At the upper limit of the
energy range, in the 1-10 eV region we note that the cross section turns upwards 
for the  $\rm 3~^3 \Sigma^- \rightarrow X~ ^3\Sigma^-$ (black line)
and the $\rm 3~^3 \Sigma^- \rightarrow A~ ^3\Pi$ (blue line), but not
for the $\rm 3~^3 \Sigma^- \rightarrow 2~^3\Pi$ (red line) process.
Referring to the energy diagram, Fig. \ref{fig1} and the moments
presented in Figs. \ref{fig2} and \ref{fig3}, we note that the classical turning point
for the entrance channel ($\rm3~^3 \Sigma^- $)  moves right from 6 a.u. to
4 a.u. over this range. This corresponds to an increasing
photon frequency (energy difference) as the kinetic energy of the
collision is converted into photon energy, 
and according to equation \ref{trate}, an enhanced decay rate.
Of course the Franck-Condon factors play an important part as well.
However, while the dipole moment increases for the
$\rm 3~^3 \Sigma^- \rightarrow X ~^3\Sigma^-$ transition,  the
$\rm 3~^3 \Sigma^- \rightarrow 2~^3\Pi$ moment drops sharply
as the turning point moves inwards which explains why the cross section for this
state continues on a downward slope in this energy region.

Radiative charge transfer has been studied for a variety of other heterogeneous ion-atom systems. 
The current total cross sections for   H$^{+}$ + S are compared with cross sections obtained 
for He$^{+}$ + O \cite{Zhao2004} and Yb$^{+}$ + Rb \cite{McLaughlin2014} in Fig. \ref{rcx-ions}.
Similarly in Fig. \ref{rate-ions} we compare the total radiative charge transfer rate coefficients, obtained by
averaging over a Maxwellian relative velocity distribution, for these systems and
 He$^{+}$ + H \cite{zygel89}, H$^{+}$ + Li \cite{Stancil1996}, and H$^{+}$ + C \cite{Stancil1998} collisions. While a clear trend is not
evident, the rate coefficients do appear to decrease with neutral atom mass for  the case of proton collisions. 
Furthermore, all rate coefficients have magnitudes between $10^{-16}$ and $10^{-13}$ cm$^3$s$^{-1}$, 
straddling the often adopted canonical value of $10^{-14}$ cm$^3$s$^{-1}$.

 In the ultra cold regime the scattering is dominated by the s-wave and an associated
complex scattering length. The sensitivity of the scattering at threshold (zero kinetic energy)  to the potential arises
from the long de Broglie wavelength and hence the long-range tail of the potential
including hyperfine structure which play an important role. Indeed the sign of the real part of
the scattering length can vary between hyperfine levels.  Our calculations for
the   inelastic (imaginary part of the phase shift) are localized, due to the dipole moment coupling,  
and hence not affected by the long-range tail of the potential. Thus we feel confident in our results shown.

Application of the optical potential approach gives the
radiative decay cross section which is the sum of radiative
charge transfer and radiative association (e.g., equations \ref{proc2}
and \ref{proc3}).  We therefore use a fully quantum mechanical
two-channel method, described for example in Ref. \cite{Stancil1996} to
compute the radiative charge transfer process directly. As
this approach is more computationally demanding, we consider
only the dominant transition and a courser grid of collision
energies. As can be seen from Figs. \ref{rcx-ions} and \ref{rate-ions}, 
the radiative charge transfer cross sections and rate coefficients are 
$\sim$3-5 times smaller than the total radiative decay implying
that radiative association (\ref{proc3}) dominates for this collision
system. As the rate coefficient for the radiative association
reaction (\ref{proc1}) via S$^+$ + H for the ground state, is five orders of magnitude smaller 
than that of reaction (\ref{proc3}) for the excited states, the formation of SH$^+$ may be dominated
by S + H$^+$ collisions which should be included in future
Sulfur astrochemical models. Explicit radiative association
calculations for process (\ref{proc3}) will be presented in a future work.

 The strength of the radiative decay rate coefficients is governed by the value of  the spontaneous 
 Einstein decay rate $\rm \Gamma(R)$ so that the transitions with the largest exothermic potential
(asymptotic separated-atom energy differences)  typically dominate.  
Fig. \ref{sh+rate} compares the current results for charge transfer due 
to non-adiabatic transfer for radial and rotational 
effects \cite{zhao2005}. Above ~1 K, radiative charge transfer will be unimportant, 
but it dominates charge transfer in the cold and ultra-cold regimes for this system.

Finally, in Figs. \ref{isotope1} and \ref{isotope2}, we illustrate the isotope effect
on radiative decay of the dominant transition for all isotopes of
Sulfur colliding with H$^+$ and D$^+$. Plotting the cross
sections as functions of the centre of mass kinetic energy
divided by the reduced mass, we find that the cross section
is insensitive to the reduced mass, at least for the considered
energy range and in regions away from the resonances.
The rate coefficients, shown in Fig. \ref{isotope2}, are nearly temperature
independent, as expected, but also reduced mass independent
except for the broad resonance feature between $\sim$3$\times 10^{-4}$
and $\sim$8$\times 10^{-3}$ K. The strength of this resonance
increases with reduced mass within each H$^+$ or D$^+$ sequence,
except for the case of $^{34}$SD$^+$, which has a rate coefficient
near 3 mK $\sim$4 times larger than all other isotopes.

 \section{Conclusions}\label{sec:Conclusions}
Radiative charge transfer for collisions of S atoms with protons has been 
studied within the optical potential method for nine electronic transitions and a large range of cold and ultracold kinetic energies. 
A fully quantal approach including radiative charge transfer is also applied.  
The lower limit for the total rate coefficients are found to be  $\sim$ 10$^{-15}$ cm$^3$s$^{-1}$ 
and virtually independent of temperature. This value is smaller than for most other 
collision systems except for He$^{+}$ + H \cite{zygel89}. 
Above $\sim$ 1 K, non-adiabatic effects in charge transfer  for H$^{+}$ + S  collisions need to be incorporated.

\ack
GS acknowledges travel support by the International Cooperation and Exchange 
Foundation of CAEP.  PCS acknowledges support from NASA grant NNX09AC46G. 
The hospitality of the University of Georgia at Athens is gratefully acknowledged by B MMcL during 
recent research visits.  B MMcL also thanks Queen's University Belfast for the award of a Visiting Research Fellowship (VRF).  
PCS and BMMcL thank the US National Science Foundation under the visitors program through a grant to ITAMP
at the Harvard-Smithsonian Center for Astrophysics.   Grants of computational time at the National Energy 
Research Scientific Computing Center in Oakland, CA, USA and at the High Performance Computing Center Stuttgart
(HLRS) of the University of Stuttgart, Stuttgart, Germany are gratefully acknowledged.
%
%
%
%
\section*{References}
\bibliographystyle{iopart-num}
\bibliography{shplus}

\end{document}